\documentclass{eas}
\usepackage{graphicx}
\newcommand{\farcs}{\mbox{\ensuremath{.\!\!^{\prime\prime}}}}% 

\begin{document}

%%-----------------------------
%%      the top matter
%%-----------------------------
\title{Characterizing Binary Properties of $5\, M_\odot$ Stars: New
  Approaches Using Cepheids } 
\author{Nancy Remage Evans}\address{SAO, MS 4, 60 Garden St.,
  Cambridge MA 02138, USA}
%\author{...}\address{...}
%\author{...}\address{...}
%
%
\begin{abstract}

Cepheids provide approaches to determining binary parameters which are
often complementary to those for main sequence massive and
intermediate mass stars.  Specifically, we are using high resolution
imaging, radial velocities, and X-ray studies to determine binary
characteristics. Among the results are that they have both a high
frequency of binary systems, and also a high proportion of triple systems.

\end{abstract}
\maketitle
%%-----------------------------
%%      your text
%%-----------------------------

\section{Introduction}
%Leuven  binary proceedings

%  $5\, M_\odot$

Great progress has been made in recent years in determining the
binary/multiple properties of massive stars.  This is important input
into our understanding of star formation.  The recent finding (Sana,
et al.  2012) that over 70\% of O stars interact with binary companions
also demonstrates the importance of binary systems in determining
the course of evolution and ultimate outcome of massive stars.  

Cepheids (typically $5\, M_\odot$) begin life as B stars.  They have
characteristics which are complementary to the more massive O stars in
ascertaining their binary properties.  In particular they have
sharp-lined spectra and hence accurate velocities.  We can also study
uncontaminated spectra for both the primary and the secondary (the
Cepheid in the visual region, a hot companion in the ultraviolet).
There is a limitation, however,  in obtaining binary properties.
Cepheids are post-red giant branch stars, and hence binaries with
periods shorter than 1 year have undergone Roche-lobe overflow, and
are no longer found in the sample of Cepheid binaries. 
We have undertaken a threefold program to determine the properties of
Cepheid binaries, as discussed in the three subsequent sections.

\section{High Resolution Studies}  We have made a snapshot survey of
69 Cepheids with the Hubble Space Telescope (HST) Wide Field Camera 3
(WFC3) in two filters which transform to V and I.  This survey can be
used in several ways.

\subsection{High Mass Companions}  We  surveyed 75 bright Cepheids
with  the International Ultraviolet Explorer satellite (IUE; Evans, 1992)  .
From this study  we can create a list of all Cepheids with companions
more massive than   $2\, M_\odot$ (Evans, et al 2013).  Of the 18 stars with
massive companions, 12 have orbits from which we can obtain the
orbital period or separation.  Three more were resolved in the WFC3
survey.  Fig 1 shows the distribution of separations from the sample.
(Note that there is no detection bias at any separation range in this
sample.)   A comparable sample (P $<$ 1 yr, mass ratio q $>$ 0.4) has
been created from the recent Raghavan et al. (2010) sample of solar
mass stars.  Fig. 1 shows that the more massive Cepheids clearly favor
shorter orbital periods than the solar mass stars.  

 \begin{figure}
\includegraphics[width=2in]{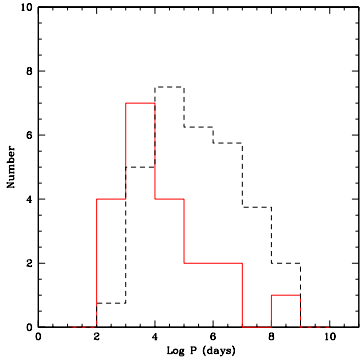}
\caption{The distribution of orbital periods.  Cepheids: solid red
  line; solar mass stars: dashed black line.  Reproduced by permission
  of the AAS from
  Evans, et al. (2013)}
\end{figure}

\subsection{Resolved Companions}  A second result from the WFC3
survey is V--(V-I) color magnitude diagrams (CMD) for the field of each
Cepheid (Fig. 2 for S Mus).  A Zero Age Main  Sequence (ZAMS) at the
distance and reddening of the Cepheid is superimposed on the CMD.  A
list is made of all possible resolved companions (down to M0).  In
order to confirm that the possible companions are physically related
to the Cepheid, we have made XMM-Newton X-ray observations of a number
of the candidates, since a physical companion of a Cepheid will be
young and much more X-ray active than a field star in a chance
alignment.  The XMM observation of S Mus has an X-ray source,
confirming that the companion is a physical companion.  Preliminary
indications, however, for several other Cepheids are that the possible
companions are chance alignments with field stars.  

 \begin{figure}
\includegraphics[width=2in]{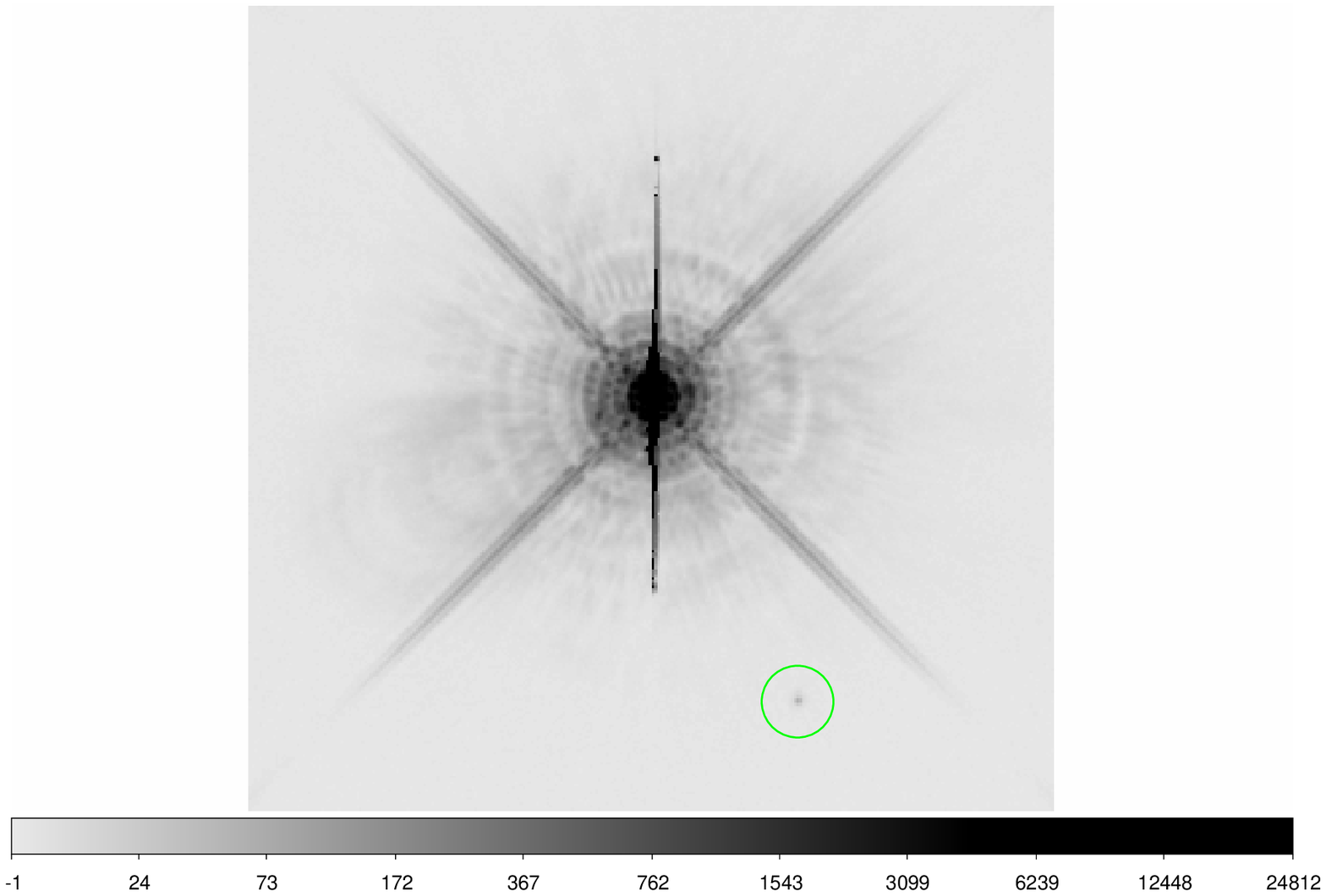}
\includegraphics[width=2in]{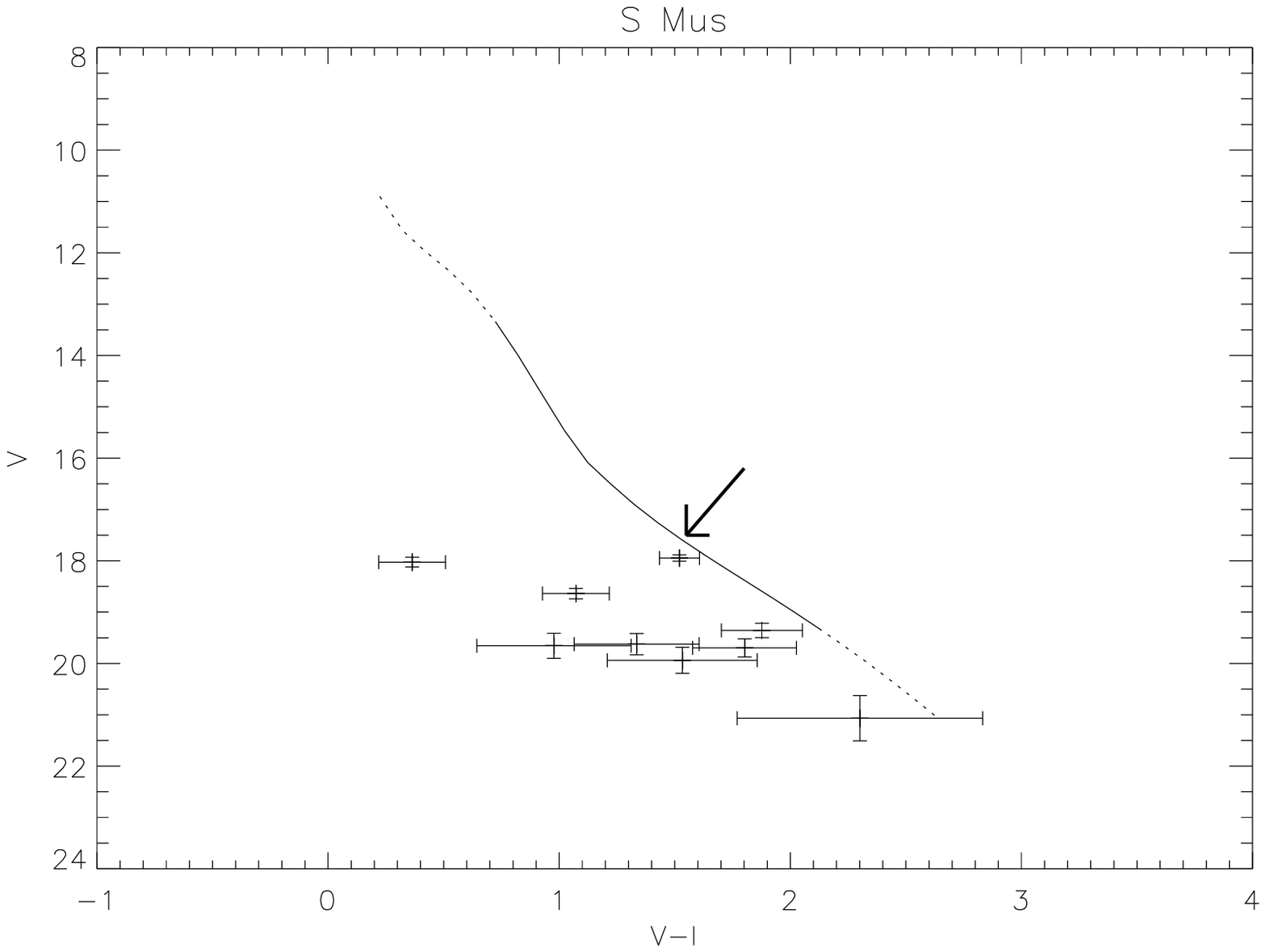}
\caption{Left: The HST WFC3 image of S Mus.  The possible companion is
circled. Right: The color magnitude diagram of the S Mus field. The
line is the ZAMS at the distance and with the reddening of the
Cepheid. The arrow indicates the possible companion.}
\end{figure}

\section{Late B Stars in Tr 16}  Late B stars--similar in mass to
Cepheids--do not in general produce X-rays.  Therefore, an X-ray
source at their location is taken to be produced by a low mass
companion.  We (Evans, et al. 2011) have identified the locations of
late B stars in a Chandra ACIS image of the cluster Tr 16 (Fig. 3 left).
The  fraction of X-ray detections indicates that  39\%
of late B stars which have a low mass companion between 1.4 and $0.5\, M_\odot$.

 \begin{figure}
\includegraphics[width=2in]{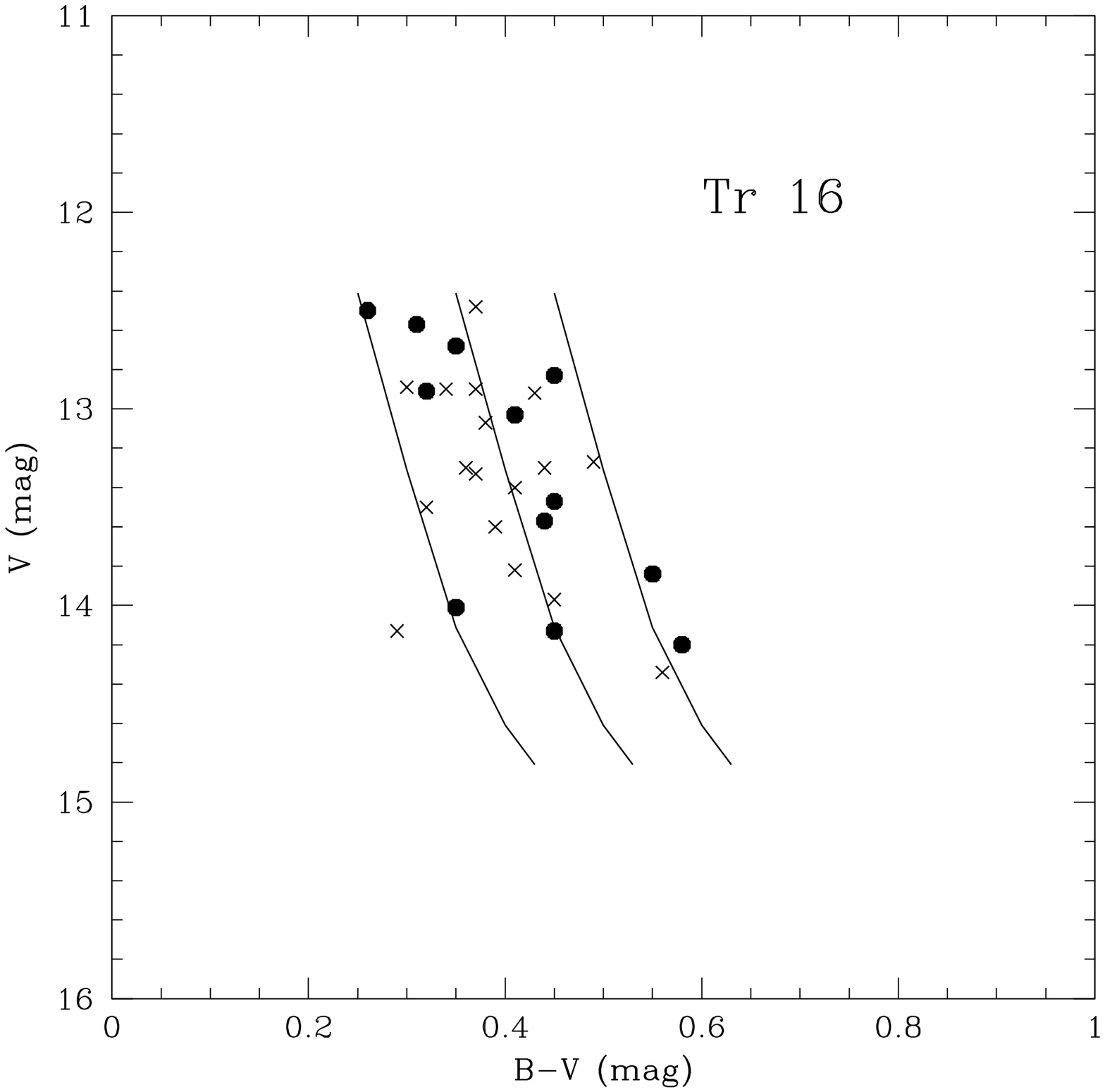}
\includegraphics[width=2in]{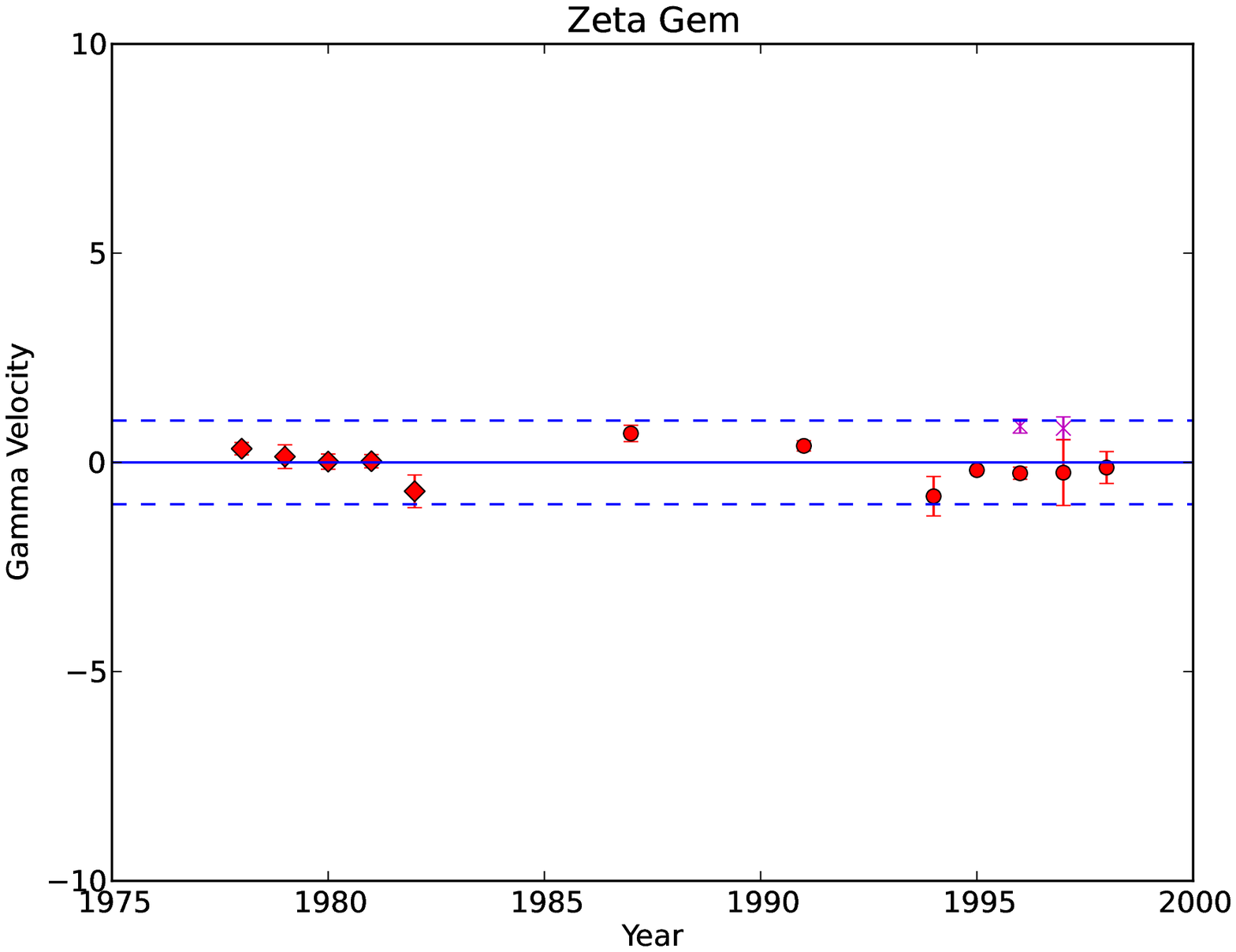}
\caption{Left: The color--magnitude diagram for late B stars in Tr 16
  (Cudworth sample).  The  lines show the ZAMS 
  with the distance and reddening of the cluster and 
 a spread of $\pm$ 0.1 mag in E(B-V).  The dots
  are detected in X-rays;  the x's are not.
 Reproduced by permission of the AAS from Evans, et al. (2011).
Right:  Annual means of radial velocity data for $\zeta$ Gem
  corrected for the pulsation velocity for using a Fourier series.
  The dashed lines indicate $\pm$ 1 km/s.  Different symbols are used
  for different data sources.} 
\end{figure}

\section{Radial Velocities}  Radial velocities accurate to 1
km/sec have been obtained in large numbers for Cepheids since the
advent of the CORAVEL radial velocity spectrometer, and subsequently
by the Moscow University group (Gorynya, et al. 1998, and
references therein).  
This means for some stars  more than 30 years  of data exist.
We have formed annual means of this data after subtracting the
pulsation curve using a Fourier series.  Fig. 3 (right) shows an
example.  We expect to be able to identity stars with an orbital
velocity amplitude $>$ 1 km/s.  Simple estimates show that with this
data we will detect 97\% of systems with the mass ratio 
q = 0.3 and P = 30 years and
77\% of systems with q = 0.1 and 30 years.  Preliminary estimates of
the results for data series up to 20 years show that about 35\% of the
Cepheids are in spectroscopic binaries.

% \begin{figure}
%% \includegraphics[key=val,key=val,...]{file}
%\includegraphics[width=2in]{zeta_gem_md_plot.ps}
%\caption{Annual means of radial velocity data for $\zeta$ Gem
%  corrected for the pulsation velocity for using a Fourier series.
%  The dashed lines indicate $\pm$ 1 km/s.  Different symbols are used
%  for different data sources.}
%\end{figure}

\section{Summary}

The combination of studies described here is extending the coverage
of q and separation/period in Cepheid binaries.  Specifically, both
spectroscopic binaries and resolved companions will be surveyed down
to mass ratios q = 0.1.  Ultimately companions will be detected at
least as close as $0\farcs5$, and probably closer.  So far the studies
have found several characteristics of binary/multiple systems.  The
frequency of triples among well-studied binaries is at least 44\%.
High mass companions (q $>$ 0.4) have smaller separations than a
comparable sample of solar mass stars.  Finally, the estimate of the
fraction of high mass companions from an IUE survey is 34\%.  The
fraction of low mass companions of late B stars is 39\%.  These
fractions should be roughly additive, implying that at least three
quarters of Cepheids occur in binary systems.   

Acknowledgments: We are grateful for financial support from HST grant
GO-12215.01-A and the Chandra X-ray Center NASA Contract NAS8-03060.

%%-----------------------------
%%      your bibliography
%%-----------------------------

\end{document}